\documentclass[twoside,twocolumn]{article}

\usepackage[sc]{mathpazo} 
\usepackage[T1]{fontenc} 
\usepackage{microtype} 
\usepackage{graphicx}
\usepackage{amsmath,amssymb}
\usepackage{multirow}
\usepackage{placeins}

\usepackage[english]{babel}
\usepackage{url}
\usepackage{multirow}
\usepackage{soul}
\usepackage[caption=false]{subfig}
\newcommand{\minitab}[2][l]{\begin{tabular}{#1}#2\end{tabular}}

\usepackage[english]{babel} 

\usepackage[margin=1in,top=1in,columnsep=20pt]{geometry} 

\usepackage{booktabs} 

\usepackage{enumitem} 
\setlist[itemize]{noitemsep} 

\usepackage{abstract} 

\usepackage{titlesec} 
\renewcommand\thesection{\Roman{section}} 

\titleformat{\section}[block]{\large\scshape\centering}{\thesection.}{1em}{} 
\titleformat{\subsection}[block]{\large}{\thesubsection.}{1em}{} 

\usepackage{fancyhdr} 
\usepackage[numbers]{natbib}

\usepackage{titling} 

\usepackage[pdfborder={0 0 0}]{hyperref} 

\pretitle{\begin{center}\huge\bfseries} 
	\posttitle{\end{center}} 
\title{EEG Connectivity Analysis Using Denoising Autoencoders for the Detection of Dyslexia} 
\author{Francisco Jesus Martinez-Murcia$^1$\thanks{Corresponding author: Francisco Jesus Martinez-Murcia. C/ Periodista Rafael Gomez 2, D1-5. 18071 Granada (Spain). Tlfn. +34 958 241717. Email: \texttt{fjesusmartinez@ugr.es}},
Andr\'es Ortiz$^1$\\
Juan Manuel G\'orriz$^2$,
	Javier Ram\'irez$^2$, \\
	Pedro Javier Lopez-Perez$^3$, Miguel López-Zamora$^3$, Juan Luis Luque$^3$ \\ \\
	$^1$Dpt. of Communications Engineering\\
	29071 University of M\'alaga, Spain\\
	$^2$Dept. of Signal Theory, Networking and Communications\\
	18071 University of Granada, Spain\\
	$^3$Dpt. Evolutive Psychology and Education\\
	29071 University of M\'alaga, Spain}
\date{28-05-2020} 
\renewcommand{\maketitlehookd}{%

\begin{abstract}
	The Temporal Sampling Framework (TSF) theorizes that the characteristic phonological difficulties of dyslexia are caused by an atypical oscillatory sampling at one or more temporal rates. The LEEDUCA study conducted a series of Electroencephalography (EEG) experiments on children listening to amplitude modulated (AM) noise with slow-rythmic prosodic (0.5-1 Hz), syllabic (4-8 Hz) or the phoneme (12-40 Hz) rates, aimed at detecting differences in perception of oscillatory sampling that could be associated with dyslexia. The purpose of this work is to check whether these differences exist and how they are related to children's performance in different language and cognitive tasks commonly used to detect dyslexia. To this purpose, temporal and spectral inter-channel EEG connectivity was estimated, and a denoising autoencoder (DAE) was trained to learn a low-dimensional representation of the connectivity matrices. This representation was studied via correlation and classification analysis, which revealed ability in detecting dyslexic subjects with an accuracy higher than 0.8, and balanced accuracy around 0.7. Some features of the DAE representation were significantly correlated ($p<0.005$) with children's performance in language and cognitive tasks of the phonological hypothesis category such as phonological awareness and rapid symbolic naming, as well as reading efficiency and reading comprehension. Finally, a deeper analysis of the adjacency matrix revealed a reduced bilateral connection between electrodes of the temporal lobe (roughly the primary auditory cortex) in DD subjects, as well as an increased connectivity of the F7 electrode, placed roughly on Broca's area. These results pave the way for a complementary assessment of dyslexia using more objective methodologies such as EEG. 
\end{abstract}
}

\begin{document}

\maketitle 


\section{Introduction}
The Developmental Dyslexia (DD) is a learning disability that hinders the acquisition of reading skills. Unrelated to mental age or inadequate schooling, it can affect between 5\% and 12\% of the population, depending on the test battery used \cite{Peterson2012}. It is characterized by from mild to severe difficulties in reading, unreadable handwriting, letter migration and common misspellings\cite{Peterson2012}. There is huge consensus \cite{cortiellacandaceStateLearningDisabilities} in that it can be a significant factor in school failure, in addition to having a harmful impact on children's self-esteem. 

The diagnosis of DD is mostly based on behavioral tests that measure reading and writing efficiency. However, the tests are often affected by exogenous variants such as children's motivation or mood, accounting for fundamental errors in the diagnosis. To overcome this problem, the standardized criteria of the 5th edition of the Diagnostic and Statistical Manual of Mental Disorders (DSM-V)\cite{american2013diagnostic} are grounded in psychometrics, but also specifies the collaboration between educators, clinicians and parents, providing other types of historical information that complement this characterization. Even in this case, the different assessments are especially designed for readers, limiting the minimum age for an early diagnosis, which may be of fundamental impact to leverage the intellectual and personal development of affected children\cite{Thompson2015}. Therefore, new objective markers to inform a more precise and early diagnosis are a paramount need. 

In this regard, many functional brain data techniques have been key in neuroscience, among others functional Magnetic Resonance Imaging (fMRI)\cite{clark2014neuroanatomical,cheng2018principal}, Magnetoencephalography (MEG) or, more recently, functional Near-Infrared Spectroscopy (fNIRS). They provide useful insight into the brain function, allowing to explore the neural basis of many disorders and diseases. Among them, Electroencephalography (EEG) is perhaps the most widespread, cost-efficient cortical brain activity detector, with the higher temporal resolution. It has countless applications ranging from human-computer interaction\cite{blankertzBCICompetition20032004,corsi2019integrating} to diagnosis, and has been extensively tested over the years\cite{ansari2019neonatal,ma2019using,li2019dynamic} in diseases and disorders such as Alzheimer's Disease (AD)\cite{Locatelli1998,ibanez2019differential,koutlis2019identification}, Parkinson's Disease\cite{klassenQuantitativeEEGPredictive2011}, Epilepsy\cite{schaper2019single,sun2019epileptic}, Stress\cite{martinez2019multiscale} or Schizophrenia\cite{akarAnalysisComplexityMeasures2015}. 

Recently, many works point to possible biological underpinnings of DD.  New models suggest that dyslexia is originated in the atypical dominant neural entrainment in the right hemisphere, strongly relying on three major rhythm categories: slow-rhythmic prosodic (0.5-1 Hz), syllabic (4-8 Hz) or the phoneme (12-40 Hz)\cite{clark2014neuroanatomical,Flanagan2018,Diliberto2018}. Among them, a Temporal Sampling Framework (TSF) for causes of DD was recently proposed\cite{kimppaImpairedNeuralMechanism2018,goswamiNeuralOscillationsPerspective2019, goswamiSpeechRhythmLanguage2019}. This hypothesis states that atypical oscillatory sampling at one or more temporal rates in children with dyslexia could cause phonological difficulties in specifying linguistic units such as syllables or phonemes. The TSF claims that atypical oscillatory entrainment at relevant rates of amplitude modulation could be one neural cause of the ``phonological deficit'' found in children and adults with dyslexia across languages and orthographies\cite{goswamiNeuralOscillationsPerspective2019}. 

During many years, it was believed that there was no relationship between EEG and DD, and just a few studies tried to shed light on the subject with mixed results\cite{boddyEvokedPotentialsDynamics1981,EEGReadingDisability}. However, research in recent years using novel spectral analysis techniques\cite{Flanagan2018,Diliberto2018} has shown that there may actually be information in the EEG signals that could be used for a biologically based diagnosis of DD. 
However, extracting meaningful data from EEG is not trivial. EEG's low signal-to-noise ratio (SNR) causes preprocessing to play an important role in the subsequent analysis. Usually, preprocessing pipelines comprise procedures that start with noise and artifact reduction, including signal averaging or Independent Component Analysis (ICA)\cite{delorme2004eeglab}. The next step is often the extraction of descriptors from the data. Spectral analysis --e.g., computing the total or average Power Spectral Density (PSD) per EEG band-- is one of the most popular\cite{Lafuente2017}, followed by computing temporal or spectral inter-channel measures\cite{roccaHumanBrainDistinctiveness2014}. 

\begin{figure*}[htp]
	\begin{center}
		\includegraphics[width=0.86\textwidth]{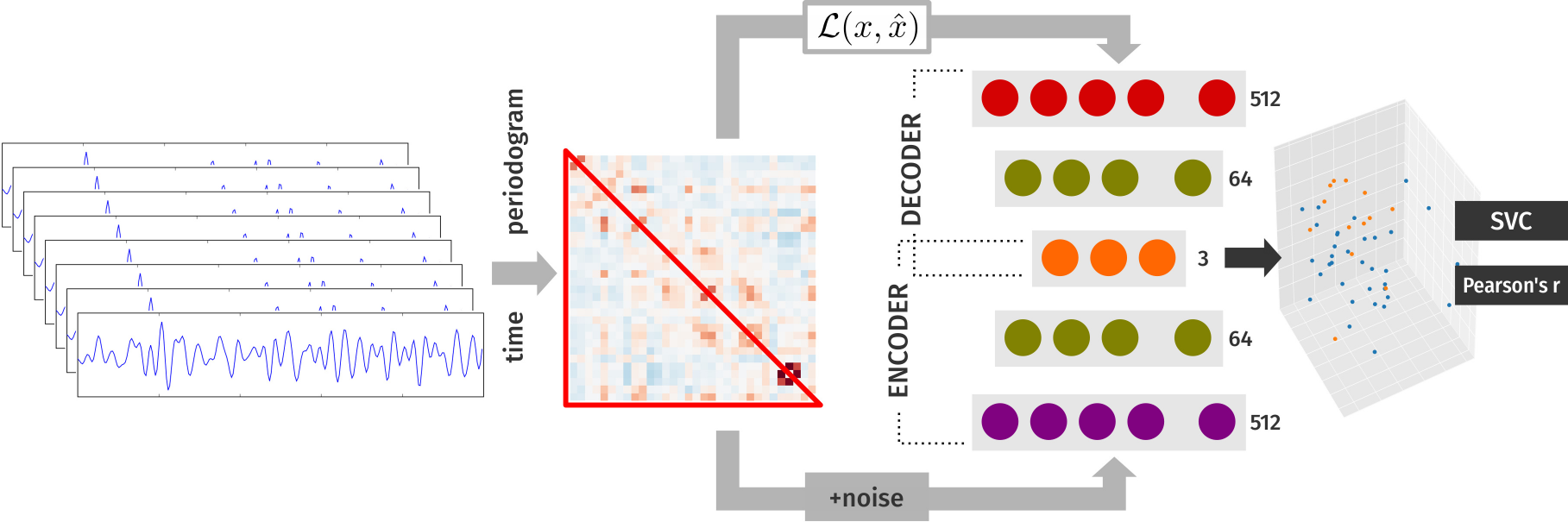}
		\caption{Schema of the proposed methodology, including how the connectivity features are obtained from time and periodogram of the EEG signals, and how this connectivity is used to train the autoencoder. The resulting features at the three-dimensional Z-layer (last layer of the encoder, first of the decoder) may be used for classification, regression or visualization.}
		\label{fig:diagram}
	\end{center}
\end{figure*}

In this context, connectivity analysis has been a major breakthrough in neuroscience\cite{Nielsen2014,Yuvaraj2016,schmidt2018tracking,galvez2018short,alarcon2018characterizing}. In the field, connectivity stands for any kind of measures that link two signals acquired at different channels, e.g. covariance or correlation. It has been demonstrated that the co-variances between signals at different regions of the brain are indicative of the underlying neural circuitry, which supports the modelling of the brain as a hyper-connected network. fMRI brain connectivity is a very common technique with outstanding performance in many diagnosis applications\cite{gavrilescu2010reduced,Ortiz2017,schmidt2018tracking,galvez2018short,alarcon2018characterizing,cheng2018principal,shin2018performance}. EEG's, however, is much less known\cite{Schoffelen2009,Sakkalis2011,ahmadlou2011functional,Lafuente2017,ahmadlou2012visibility,ahmadlou2012fuzzy,ahmadlou2013spatiotemporal,ahmadlou2014complexity,ahmadlou2017complexity,deletoile2017graph,tapani2019time}. Nevertheless, EEG connectivity revealed itself as a very promising technique in Delignani et al.\cite{deligianniRelatingRestingstateFMRI2014}, where the authors already used spectral information to compute the adjacency matrices, and could even predict fMRI connectivity. Spectral connectivity was also successfully used in Martinez-Murcia et al.\cite{martinez-murciaPeriodogramConnectivityEEG2019} for the diagnosis of DD. However, spectral processing is mainly used to filter out frequency bands (alpha, beta, delta, theta) and, with the exception of spectral coherence\cite{roccaHumanBrainDistinctiveness2014}, the periodogram has hardly been used for direct connectivity estimation. 

These matrices can be directly used as markers for diagnosis. However, the variability inherited from low SNR signals, and the small sample size problem\cite{Duin00}, very frequent in these experimental setups, can affect the results. Therefore, new methodologies to reduce the dimensionality of the data can be of great help, both to avoid the small sample size problem and to reveal their underlying structure in the data via manifold learning\cite{ChaSchZie06}. In the first case, some works explore algorithms such as Principal Component Analysis (PCA)\cite{martinez-murciaPeriodogramConnectivityEEG2019} to reduce the feature space. On the other hand, manifold learning has also been used for feature extraction while at the same time, characterizing the underlying structure of biomedical signals in many studies\cite{martinez-murciaStudyingManifoldStructure2019}. 
Manifold learning stands for a geometric interpretation of non-linear decomposition methods, following the manifold assumption\cite{ChaSchZie06}: that natural high dimensional data concentrates close to a non-linear low-dimensional manifold. 
Its advantages are a more powerful non-linear modelling while providing better features for classification and visualization.
Many manifold learning approaches can be found in the literature, among them those based on minimum distance\cite{hettiarachchi2015multi} or locally linear embedding\cite{saul2003think}. However, most of them have been outperformed in latter years by neural network architectures. In this context autoencoders, a self-supervised encoder-decoder architecture, have been widely applied\cite{martinez-murciaStudyingManifoldStructure2019}. In particular, denoising autoencoders\cite{vincent2010stacked} improve the representation of the information by adding noise to the input and training on the loss between the noisy and the real input, learning how to separate useful information from noise. 
Once trained, the model can map a corrupted example back to an uncorrupted one, and the encoder part can be used to project from the high-dimensional space of the adjacency matrices to just a few coordinates over a collinear manifold that may be representative of each data point\cite{vincent2010stacked}. 

Here, we test the TSF's atypical oscillatory sampling hypothesis by studying EEG signals collected by the LEEDUCA Project\cite{alvarez2017syllabic} in southern Spain to analyze whether and how EEG connectivity is affected by DD. To do so, autoencoders will be used to inform a self-supervised decomposition of EEG's inter-channel temporal and spectral connectivity, as shown in Figure~\ref{fig:diagram}. The resulting manifold will be studied for validity by checking its correlation to reading and writing skill performance evaluations done by researchers\cite{de2017longitudinal}, and by obtaining the classification performance in a binary (DD vs control) approach. The whole LEEDUCA dataset as well as the EEG cohort are presented at Section~\ref{sec:leeduca}, followed by a description of how the spectrum is computed (Section~\ref{sec:spectrum}) the 
connectivity measures (Sec.~\ref{sec:connectivity}) and the denoising autoencoder (Sec.~\ref{sec:autoencoders}). Finally, at Sections~\ref{sec:results} and \ref{sec:discussion} the results of applying this methodology will be presented and discussed.

\section{Materials and Methods}

\subsection{EEG Dataset}\label{sec:leeduca}
\subsubsection{LEEDUCA cohort}
The LEEDUCA study is a longitudinal study aimed at assessing specific learning difficulties and their evolution during infancy \cite{alvarez2017syllabic}. It follows a large cohort ($N \approx 700$) of students at 30 schools in southern Spain from five to eight years. Students undergo a complete battery of cognitive and linguistic tasks applied by expert psychologists, whose content is based on recent studies \cite{de2017longitudinal}, including several tasks. Among them, Phonological Memory (PM) and Phonological Awareness (PA) count the number of suppressed syllables or phonemes in a variety of listening tests; Reading Efficiency (RE), which measures the efficiency (words per minute) in identifying two, three and four syllable words and pseudowords. Reading Speed (RS) measures the number of words per minute (efficiency) on reading a real text, Reading Comprehension (RC) presents a text and a questionnaire, and counts the number of correct answers and Rapid Symbolic Naming (RSN) measures the time on a rapid automatized naming of some object presented. This study was approved by the Medical Ethical Committee of the University of M\'alaga (05/02/2020 PND016/2020), according to the dispositions of the World Medical Association Declaration of Helsinki. It was also supported by the University of Malaga (infrastructure project UNMA15-CE-3657) and the Education Office of the regional government of Andalusia (Spain), which granted our researchers permission to carry out the study in different public schools, and it was approved, funded and supervised by the Spanish Ministry of Science within the framework of the national project PSI2015-65848-R. Labeling for this paper was derived from a complete report at age seven, received by the Special Education School Services (SESS) that coordinate the project. Individuals meeting specific criteria of the Diagnostic and statistical manual of mental disorder DSM-V (a reading performance 1.5 standard deviations below the mean for age 7) were labeled as DD\cite{american2013diagnostic}, and the rest were considered as subjects of the control group (CN). 

\begin{table}
	\caption{Demographics and test results of the EEG cohort. Units for each category are provided in parentheses ($n$: number of correct answers, $t$: time in seconds, efi: efficiency, measured in items per minute), and the standard deviation per group is provided in brackets.} \label{tab:demo_eeg}
	\begin{tabular}{lrr}
			\toprule
			Group & Control & DD\\ 
			\midrule
			$N$			&     32			& 16			\\
			Age	(months)	  & 94.1  {[}3.3{]}				& 95.6 {[}2.9{]}\\
			PM ($n$)        &      16.382 [5.551] &      13.219 [3.843]\\
			PA ($n$)         &      11.206 [3.210] &       9.094 [3.489]\\
			RE-words (efi)   &      17.152 [2.984] &      10.730 [3.780]\\
			RE-pseudo2$^1$ (efi) &      35.633 [8.187] &      21.419 [5.896]\\
			RE-pseudo3$^2$ (efi) &      24.342 [6.127] &      16.394 [6.242]\\
			Reading Speed (efi)      &      15.710 [3.08] &       9.206 [3.550] \\
			RSN ($t$)&      18.969 [5.999] &      21.335 [6.767]\\ 
			RC ($t$)        &       8.106 [1.339] &       7.625 [1.443]\\
			\bottomrule\\
			
			\multicolumn{3}{p{8cm}}{$^1$2-syllable pseudo-words, $^2$3-syllable pseudo-words.}
	\end{tabular}
\end{table}

A subset of the LEEDUCA study were selected by the Special Educational Need Services of the regional school system to test the TSF for dyslexia \cite{kimppaImpairedNeuralMechanism2018,goswamiNeuralOscillationsPerspective2019, goswamiSpeechRhythmLanguage2019} using EEG. This cohort  ($N=48$) included 32 skilled readers (17 males) and 16 dyslexic readers (7 males) matched in age ($t = -1.4, p=0.180$). Details of demographics as well as some test outcomes for the EEG cohort can be found at Table~\ref{tab:demo_eeg}. Specific written permission was asked for this experiment and subjects came accompanied by their parents. They underwent 5-minute sessions while presenting a rhythm-modulated auditive stimulus, in order to identify if there exists an abnormal neural processing of speech envelope modulation rates in subjects with DD. This stimulus consisted of Amplitude-Modulated (AM) white-noise at a fixed rate of 2, 8 and 20 Hz, which correspond to stress word patterns, syllable Spanish rate and phoneme segmentation, as in the work of De Vos et al.\cite{de2017longitudinal}

\begin{figure}
	\begin{center}
		\includegraphics[width=0.32\textwidth]{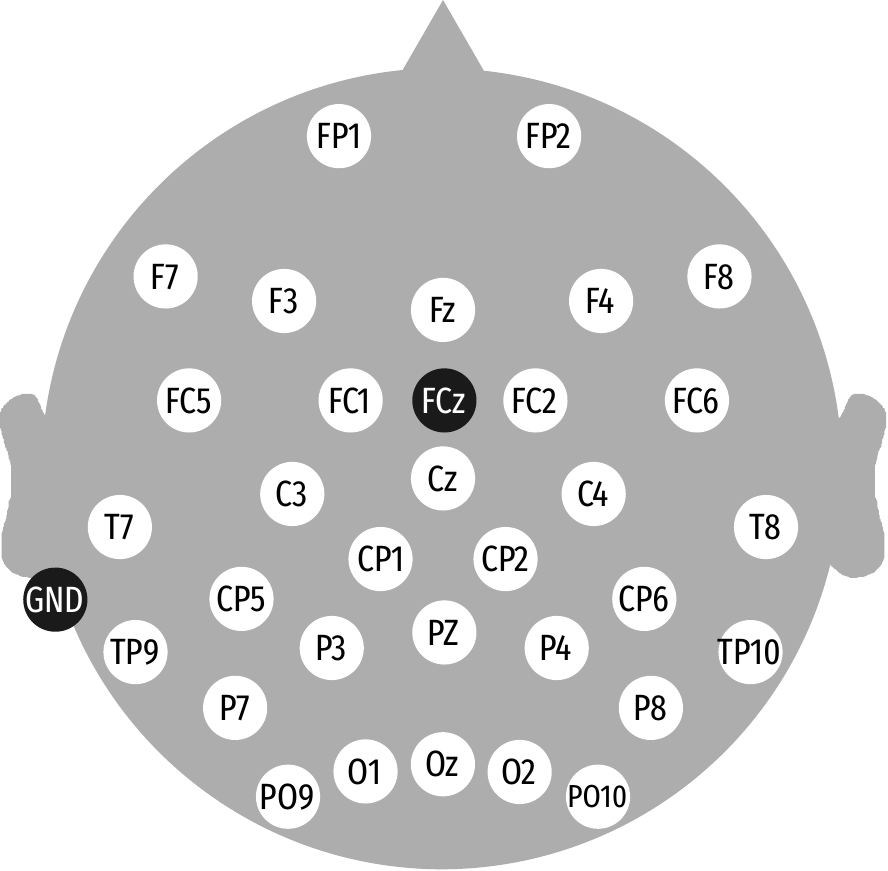}
		\caption{EEG electrode placement according to the 10-20 coordinate system, including the ground and reference electrodes.}
		\label{fig:eeglayout}
	\end{center}
\end{figure}

\subsubsection{Signal Acquisition and Preprocessing}
EEG was recorded using the BrainVision actiCHamp Plus amplifier with actiCAP snap high impedance active electrodes in a 32-channel 10-20 standard layout (see Fig.~\ref{fig:eeglayout}) plus one reference (REF) at FCz and one ground (GND) electrode attached to the mastoid bones (behind the ears). The equipment was powered by Li-ion batteries to ensure isolation from power line and reducing the noise, and signals were recorded at a sampling rate $f_s=500Hz$ and band-pass filtered from the DC component to 140 Hz. 
EEG data were preprocessed using the BrainVision analyzer software. Electro-Ocular (EOG) artifacts were removed using the ocular correction implemented in the software, that detects and marks artifacts with a Mean Slope algorithm\cite{gratton1983new}, and then eliminates the contribution of the components to the Global Field Power during blink intervals. 
Afterwards, data was filtered out for power line interference (notch filter at 50 Hz), and the signals were cropped into five-second segments. EEGLAB’s\cite{delorme2004eeglab} has been used under expert supervision for the detection of potential abnormal values, abnormal trends, improbable data, abnormal distribution and abnormal spectra with default parameters. Segments without potential artifacts were kept, and those with potential artifacts were examined by an expert neurophysiologist to either keep or reject the segment in further analyses.

\subsection{Spectral Estimation}\label{sec:spectrum}
In order to estimate the spectral connectivity, we cannot rely on average estimates of the Power Spectral Density (PSD) of different bands and channels; the whole spectrum is needed. And for this purpose, each subject's 32-channel five-second segments is used to estimate the periodogram. For the first approach (noted `pgram'), the raw periodogram is obtained from the discrete Fourier's transform (DFT) of each channel and segment. On the other hand, the second approach (noted `welch') uses the Welch's periodogram estimation method\cite{Welch1967}. In this approach, the signal is first divided on several subsegments of length 500 with an overlapping of 250 samples. Then, the segments are windowed (using the `hanning' window) and the periodogram is estimated from the windowed segments using the DFT. Finally, the periodograms of the subsegments are averaged to produce an estimate of the spectrum less affected by the noise, at the cost of reducing the spectral resolution. Both periodogram estimation methods are implemented using the \texttt{scipy} package\cite{virtanen2020scipy}.

Finally, we can use directly the pgram or welch periodograms for each segment (the per-segment approach), or apply the same modified Welch's periodogram used by Martinez-Murcia et al.\cite{martinez-murciaPeriodogramConnectivityEEG2019}, in which the periodograms --regardless of the method used-- of all segments for one subject are again averaged. This is noted as the `per-subject' approach, obtaining just one spectral estimate per subject and channel. This later approach has the advantage of reducing the overall noise of the spectral estimate at both the segment and subject level, although it also reduces the spectral resolution. 

\subsection{Connectivity Features}\label{sec:connectivity}
The characterization of the brain as a network using functional information is commonplace in current brain studies\cite{ortiz2017learning,Verly2014,Yuvaraj2016,Ecker2012b,Schoffelen2009,Sakkalis2011}. This leads to the so-called connectome: a complete mapping of all connections between regions, in the form of an adjacency matrix, usually containing the covariance (and other derived measures) between fMRI signals at different regions. Similarly, the temporal covariance between EEG electrodes has also been assessed in several works\cite{Schoffelen2009}. However, the covariance is a statistic not restricted to the temporal domain, and in consequence, it could be used to quantify co-varying changes in any kind of signal, e.g. the spectrum. 

Many algorithms exist to estimate the real \textbf{covariance} $\Sigma$ of a set of measurements, such as the Ledoit-Wolf estimator\cite{Ledoit2004}. However, when dealing with connectivity, the \textbf{precision matrix} (or inverse covariance) $\Theta$ is usually of greater interest, since it accounts just for direct connections between nodes. In these cases, a robust covariance estimator is preferred. The problem here becomes to minimize: 
\begin{equation}\label{eq:maxcov}
	\log \det \Theta -\text{tr}(S\Theta) - \rho \|\Theta\|_1
\end{equation}
where $S$ is the sample covariance, tr() denotes the trace and $\|\Theta\|_1$ is the $L1$ norm of $\Sigma$ (the sum of all the absolute values). Note that the $L1$ norm  is used to enforce sparsity: the higher the parameter $\rho$ is, the sparser the final estimate will be. 

To solve this problem, Friedman et al.\cite{friedmanSparseInverseCovariance2008} show that the minimization of Eq.~(\ref{eq:maxcov}) is a convex problem, and by estimating an estimate $W$ of $\Sigma$, instead of $\Sigma^{-1}$, it was shown that the optimization problem could be applied to each row and corresponding column of $W$ as in a block coordinate descent: 
\begin{equation}
W= \begin{pmatrix}
W_{11} & w_{12}\\
w_{21}^T & w_{22}
\end{pmatrix}, \quad
S= \begin{pmatrix}
S_{11} & s_{12}\\
s_{21}^T & s_{22}
\end{pmatrix}
\end{equation}
to propose a solution of the problem based on:
\begin{equation}
	\min_\beta \{\frac{1}{2}\|W_{11}^{1/2}\beta-b\|+\rho\|\beta\|_1\}
\end{equation}
where $b=W^{-1/2}_{11} s_{12}$, which resembles a lasso regression. Then, a coordinate descent algorithm is used to iterate over $W$ and obtain the final covariance estimate. Since the model is sensitive to $\rho$ variations, an inner cross-validation was used to automatically set it. 

Once the \textbf{covariance} model estimate $W$ has been obtained, the \textbf{precision} is obtained as the pseudo-inverse of $W$. Finally, the \textbf{correlation} $R$ and \textbf{partial correlation} (PC) $T$ matrices are obtained from the covariance matrix as: 
\begin{equation}
	R = W d d^T\quad \text{and} \quad T = -\Theta g g^T
\end{equation}
where
\begin{equation}
d = \frac{1}{\sqrt{\text{diag}(S)}}
\quad \text{and} \quad g = \frac{1}{\sqrt{\text{diag}(\Theta)}}
\end{equation}

Since these four matrices are symmetrical, the values at the lower triangular part of the matrix will be selected as features. The values for connectivity have been estimated using the \texttt{scikit-learn} python package\cite{pedregosa2011scikit}.

\subsection{Denoising Autoencoder}\label{sec:autoencoders}
Autoencoders (AEs) are a self-supervised neural network that is frequently used for feature extraction\cite{Baldi2012,Ortiz2017}. It consists of a connection of an encoder and a decoder network, the former reducing the dimensionality of the input features $x$ down to a bottleneck vector $z$ of length $\text{len}(z)<<\text{len}(x)$. Then, the output of the encoder is connected to a decoder network whose only purpose is to reconstruct the original signal $\hat{x}$ using solely the information at $z$. The intermediate layer, or bottleneck, is commonly known as $z$-layer. 

The complete encoder-decoder network is trained by minimizing the reconstruction error (in our case, the Mean-Squared Error, or MSE) between the input and the output vector, in a self-supervised scheme: 
\begin{equation}
	\mathcal{L} = \frac{1}{N} \sum_i (x_i-\hat{x}_i) ^2
\end{equation}

A particular type of AE is the Denoising Autoencoder (DAE)\cite{vincent2010stacked}, in which the forward pass of the network uses a corrupted input by adding white noise $\dot{x} = x + n$. 
By adding noise to the input and backpropagating the loss between the corrupted and the clean input, the model learns how to retain useful information and discard noise, which was given the geometric interpretation under the manifold assumption: that natural high dimensional data concentrates close to a non-linear low-dimensional manifold\cite{ChaSchZie06}. Under this interpretation, the outputs of the latent neurons (neurons at the intermediate layer of an autoencoder) can be considered a set of coordinates in the latent space, to which the higher-dimensional input is projected. In this work, the noise is re-sampled from a white noise of standard distribution $\mathcal{N}(0,1)$ in every epoch, which also acts as a data augmentation approach at the same time that avoids local minima, providing robustness to overfitting. 

Our architecture is composed of a three-layer perceptron as the encoder (see Figure~\ref{fig:diagram}), with layer sizes of 512, $n_{hid}$ and 3, and the corresponding decoder of layer sizes 3, $n_{hid}$ and 512. $n_{hid}$ --the number of neurons at the hidden layer of the encoder and decoder-- was chosen by grid search in powers of two, by reporting the reconstruction loss for each $n_{hid}$ and connectivity feature. A $n_{hid}=64$ was chosen, corresponding to the value for which the validation loss stopped decreasing. The number of neurons at the Z-layer was set to 3, to be able to visualize the distribution of subjects in a three-dimensional space. Visualizing the Z-layer space improves the interpretability of our methodology, one of the main objectives of this work. The layers had activations (ELU, linear) for both the encoder and the decoder respectively. ELU stands for the Exponential Linear Unit function, a widely used activation function defined as:
\begin{equation}
	ELU(x) = \max(0,x)+\min(0,\alpha(\exp(x)-1))
\end{equation}
Batch normalization was used after each ELU activation. The whole system was trained first with Adam\cite{kingmaAdamMethodStochastic2014} and later finetuned with Stochastic Gradient Descent (lr=0.001). Early stopping was used in both cases, controlled by an independent validation subset (15\% of the data), and batch size was 16 (in per-subject connectivity) and 64 (for per-segment connectivity). The autoencoder was implemented and trained using the \texttt{pytorch} python framework\cite{paszke2019pytorch}.

\section{Results}\label{sec:results}

\begin{figure*}[t]
	\begin{center}
		\includegraphics[width=0.40\textwidth]{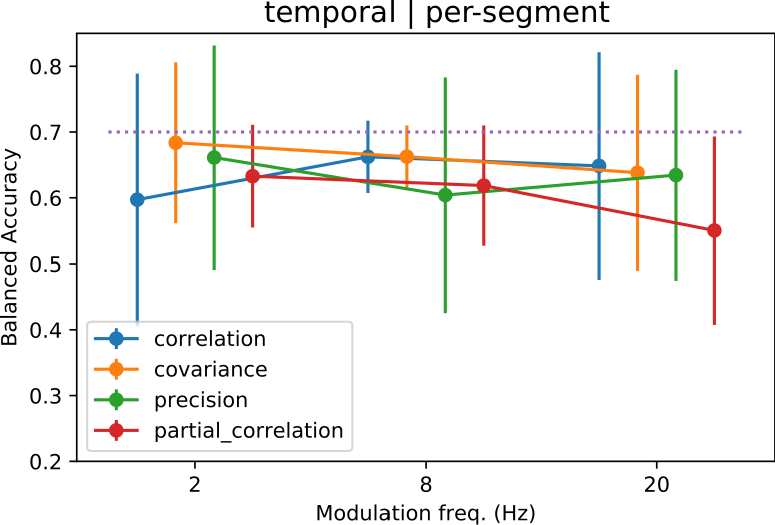}\hfil
		\includegraphics[width=0.40\textwidth]{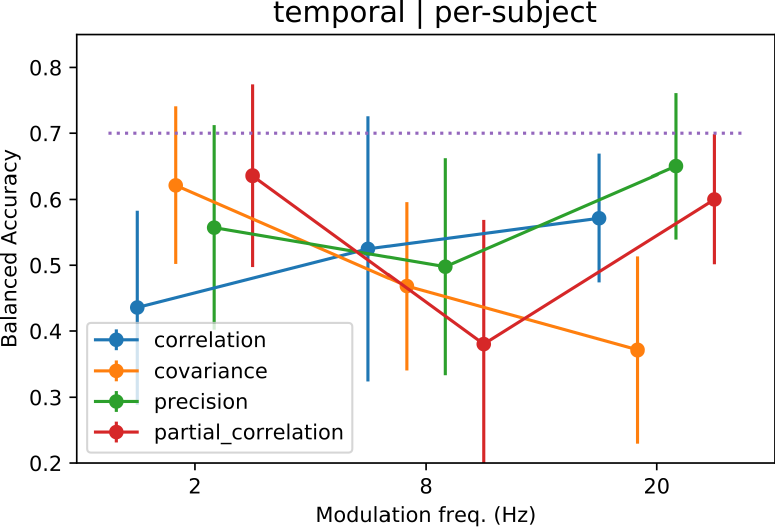}
		\caption{Performance of the different temporal connectivity measures under the per-segment (left) and per-subject (right) approaches. BA and its STD are provided for each measure (in color) and frequency modulation of the stimuli (x-axis). Note that the points are not exactly located at their x-position in order to ease visual inspection of the performance and its trend. }
		\label{fig:no_seg_grouped}
	\end{center}
\end{figure*}

\subsection{Experimental Setup and Evaluation}\label{sec:setup}
In accordance to the TSF for dyslexia \cite{kimppaImpairedNeuralMechanism2018,goswamiNeuralOscillationsPerspective2019, goswamiSpeechRhythmLanguage2019}, it would be likely that atypical oscillatory sampling patterns are found in EEG signals. The main aim of this work is therefore to test whether and how EEG connectivity is affected by DD in school-aged children. To do so, the self-supervised decomposition of temporal and spectral connectivity between EEG channels and their relationship with DD is studied within two main categories:
\begin{itemize}
	\item A \textbf{correlation} analysis. The Pearson's Correlation ('$r$') and statistical significance ($p$-value) between each Z-layer feature and the language and cognitive task performance of children of the EEG cohort (see Section~\ref{sec:leeduca}) are estimated. Its objective is twofold: to validate the autoencoder decomposition and evaluate which tasks are more related to EEG connectivity, allowing for a comparison with other works in the literature\cite{Diliberto2018}. 
	\item A \textbf{classification} analysis, in which the Z-layer features (Sec~\ref{sec:autoencoders}) are used to train and test a Support Vector Machine classifier (SVC)\cite{Vapnik1998} with Radial Basis Function (RBF) kernel ($\gamma=0.1, C=1$), using balanced class weight. The performance is estimated within a 5 fold stratified cross-validation (CV) loop\cite{Kohavi1995} using a series of performance metrics that are described below.
\end{itemize}

The performance of the classification is estimated using a number of metrics derived from the confusion matrix, formulated in relation to the number of True Positives (TP), True Negatives (TN), False Positives (FP) and False Negatives (FN):
\begin{itemize}
	\item Accuracy (acc.) and its standard deviation (STD) over all CV folds: \begin{equation}
		acc.=\frac{TP+TN}{TP+TN+FP+FN}
	\end{equation}
	\item Sensitivity (sens.) and its STD:
	\begin{equation}
		sens. = TP/(TP+FN)
	\end{equation}
	\item Specificity (spec) and its STD over CV folds:
	\begin{equation}
		spec. = TN/(TN+FP)
	\end{equation}
	\item F1-score:
	\begin{equation}
		F1 = 2TP/(2TP+FP+FN)
	\end{equation} 
	\item Balanced Accuracy (BA), especially designed for unbalanced datasets: 
	\begin{equation}
		BA = (sens.+spec.)/2
	\end{equation}
	Note that, given the unbalanced nature of our data, balanced accuracy is always preferred to regular accuracy. 
\end{itemize} 
The STD is provided because the sample size is small in the per-subject approach, and as a consequence, the variance of the performance across loop might be high. Additionally, the Receiver-Operating Characteristic (ROC) curve and the area under the curve (AUC) are provided as an additional performance measure. 

The classification and correlation analyses are performed on two main models: 
\begin{itemize}
	\item A \textbf{temporal connectivity} model, in which the adjacency matrix is estimated from the EEG multi-channel segments, in a per-segment or per-subject --average of all matrices belonging to a subject-- approach. This is presented at Section~\ref{sec:temporal_results}.
	\item A \textbf{spectral connectivity}, in which the adjacency matrices are estimated using the periodograms of different segments (see Section~\ref{sec:spectral}). Here, they are used to assess which spectral estimation methods and adjacency matrices generate a better modelling of the manifold for our purposes. This is done again in a per-segment and per-subject (connectivity of the average spectrum) approaches. 
\end{itemize}

\subsection{Temporal Connectivity Model}\label{sec:temporal_results}
In this section the results for the temporal connectivity are provided. Temporal connectivity is the standard measure in most connectivity studies\cite{ortiz2017learning,Verly2014,Yuvaraj2016,Ecker2012b}, and describes how the Blood Oxygen Level-Dependent (BOLD) or EEG signals in different regions co-vary over time. In Figure~\ref{fig:no_seg_grouped}, the classification performance (estimated by a SVC) of the AE features at the Z-layer is depicted for both the per-segment and per-subject scenarios.

\begin{table*}[t]
	\caption{Performance values for the temporal connectivity measures in DD diagnosis.} 
	\label{tab:temp_overview}
	\begin{tabular}{lrrrrrrrrrr}
			\toprule
			& $f_m$ &  acc. [STD] & sens. [STD] & spec. [STD] & F1 & BA &   AUC \\
			\midrule  
			\multirow{3}{*}{\minitab[c]{covariance\\(per-segment)}} & 2 & 0.783 [0.122] & \textbf{0.471} [0.312] & 0.896 [0.236] & \textbf{0.598} &  \textbf{0.684} & 0.663 \\
			& 8 & \textbf{0.827} [0.047] & 0.325 [0.000] & \textbf{1.000 [0.098]} & 0.491 &  0.663 & \textbf{0.762} \\
			& 20 & 0.733 [0.149] & 0.441 [0.339] & 0.836 [0.264] & 0.549 &  0.638 & 0.689 \\
			\midrule
			\multirow{3}{*}{\minitab[c]{covariance\\(per-subject)}} & 2 & \textbf{0.612 [0.119]} & \textbf{0.643 [0.208]} & 0.600 [0.307] & \textbf{0.629} &   \textbf{0.621} & 0.465 \\
			& 8 & 0.574 [0.128] & 0.231 [0.363] & 0.706 [0.324] & 0.303 &   0.468 & \textbf{0.480} \\
			& 20 & 0.531 [0.142] & 0.000 [0.000] & \textbf{0.743 [0.251]} &   - &   0.371 & 0.363 \\
			\midrule
			\multirow{3}{*}{\minitab[c]{precision\\(per-segment)}}& 2 & \textbf{0.708 [0.170]} & 0.562 [0.183] & \textbf{0.761 [0.206]} & \textbf{0.624} & \textbf{ 0.661} & \textbf{0.740} \\
			& 8 & 0.672 [0.179] & 0.466 [0.324] & 0.743 [0.303] & 0.540 &  0.604 & 0.598 \\
			& 20 & 0.667 [0.160] & \textbf{0.567 [0.231]} & 0.702 [0.275] & 0.608 &  0.635 & 0.590 \\
			\midrule
			\multirow{3}{*}{\minitab[c]{precision\\(per-subject)}}& 2 & 0.612 [0.155] & 0.429 [0.339] & 0.686 [0.313] & 0.492 &   0.557 & 0.535 \\
			& 8 & \textbf{0.617 [0.164]} & 0.231 [0.363] & \textbf{0.765 [0.303]} & 0.315 &   0.498 & 0.419 \\
			& 20 & 0.592 [0.111] & \textbf{0.786 [0.116]} & 0.514 [0.264] & \textbf{0.692} &   \textbf{0.650} & \textbf{0.618} \\
			\bottomrule 
	\end{tabular}
\end{table*}

The most obvious difference is that the performance dramatically decreases from the per-segment to the per-subject scenario. This is expected, given that it involves a significant reduction of the sample size--from 1462 adjacency matrices to 48--, increasing the variance and decreasing the performance of any classification system. Therefore, we will focus on the general trends. 

In the per-segment scenario, the performance behaves similarly across all measures, with a subtle trend to diminish at higher modulation frequencies. In the per-subject approach, however, the only measure that maintains this trend is the covariance, whereas the rest have a noticeable decrease at the 8Hz band. It is interesting to note that, whereas covariance reaches its peak at $f_m=2$ Hz, precision --the inverse covariance (see Sec.~\ref{sec:connectivity})-- does so at the highest $f_m$. We provide a deeper look at the performance of these measures at Table~\ref{tab:temp_overview}. 

\begin{figure}
	\begin{center}
		\includegraphics[width=\columnwidth]{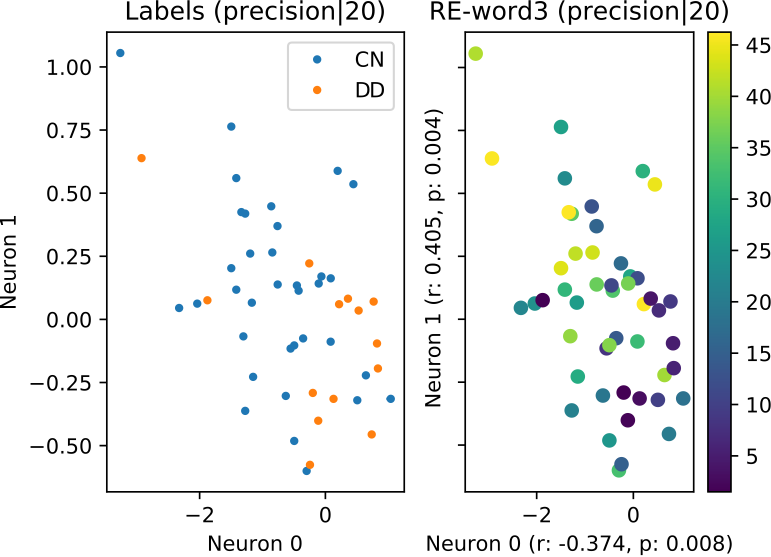}
		\caption{Manifold representation of the AE trained with temporal precision matrices obtained with 20Hz stimuli and the per-subject approach, and its relation with the labels (left, DD and controls) and one of the tasks (right, 3-syllable words RE).}
		\label{fig:no_seg_manifold}
	\end{center}
\end{figure}

Here, the differences between the per-subject and per-segment approaches vary significantly. While the per-segment approach achieves a balanced accuracy above 0.7 in both covariance and precision, the per-subject hardly gets a 0.6 for the later and less than 0.5 for the former. There is even a notorious example with 0 sensitivity for the covariance matrix at a modulation $f_m=20$Hz. In Figure~\ref{fig:no_seg_grouped}, it is shown that the covariance connectivity yielded larger performance for the 2Hz stimuli whereas the precision did at 20 Hz. This is especially evident when assessing the sensitivity and the F1-score, that are higher at $f_m=2$ for the covariance (both in per-segment and per-subject) and at $f_m=20$ for the precision. In these cases, the resulting manifold achieves a good trade-off between sensitivity and accuracy (a BA above 0.6), which may hint a link between reading difficulties and the self-supervised decomposition of the connectivity. This link is easily spotted at Figure~\ref{fig:no_seg_manifold}-left, where most DD-affected subjects appear at the right lower area of the subject distribution in the AE representation. 

\begin{figure*}[ht]
	\begin{center}
		\subfloat[Periodogram estimation, per-segment\label{fig:pgram_seg}]
		{\includegraphics[width=0.4\textwidth]{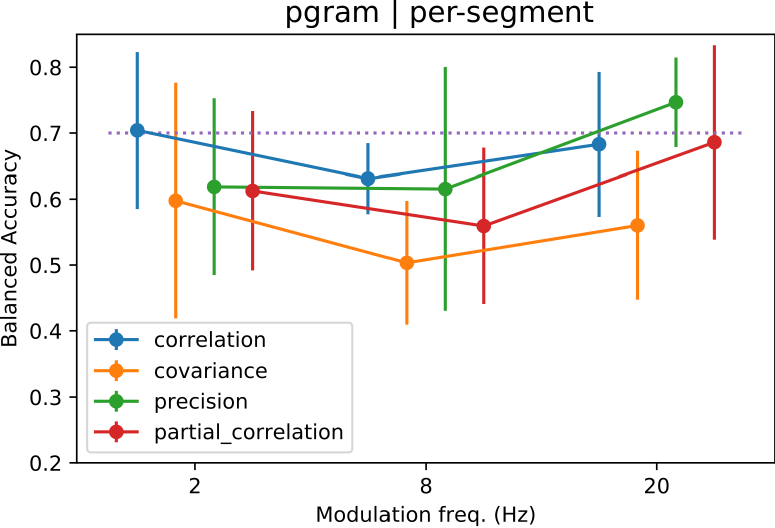}}\hfil
		\subfloat[Welch's estimation, per-segment \label{fig:welch_seg}]
		{\includegraphics[width=0.4\textwidth]{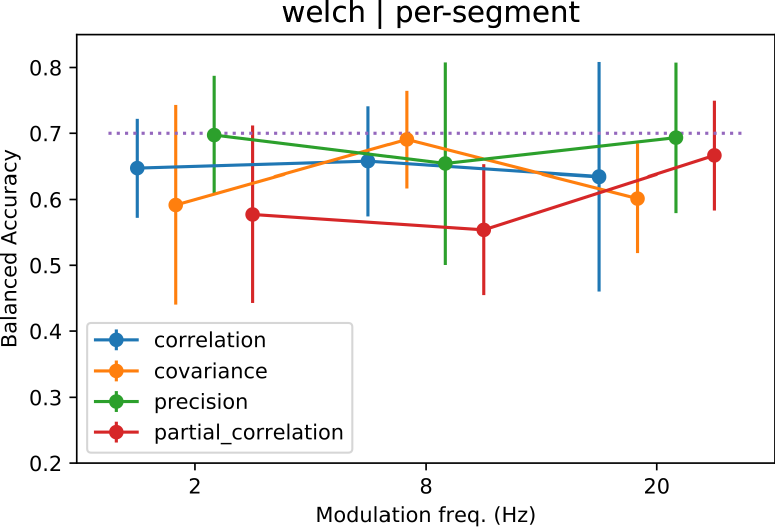}}\\
		\subfloat[Periodogram estimation, per-subject \label{fig:pgram_sub}]
		{\includegraphics[width=0.4\textwidth]{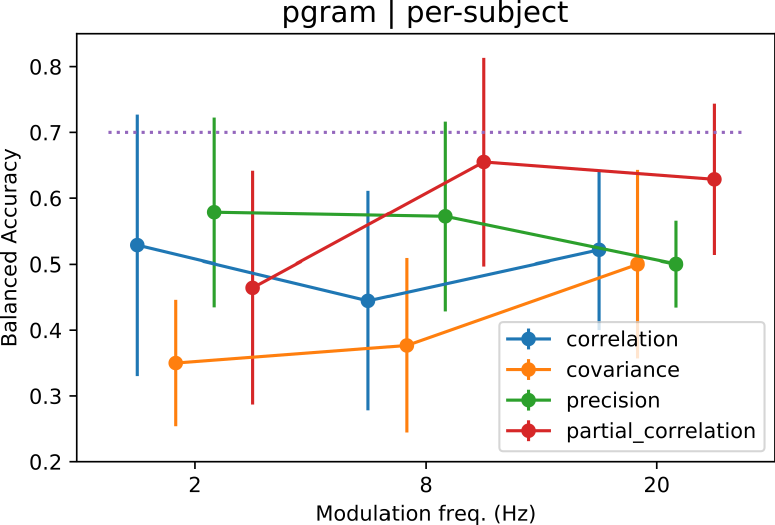}}\hfil
		\subfloat[Welch's estimation, per-subject \label{fig:welch_sub}]
		{\includegraphics[width=0.4\textwidth]{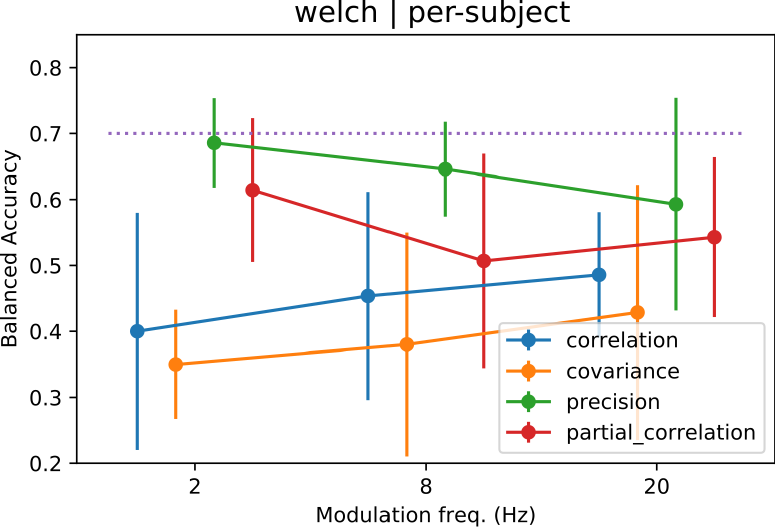}}
		\caption{Performance (BA and STD) of the different periodogram connectivity measures for each modulation frequency.}
		\label{fig:seg_grouped}
	\end{center}
\end{figure*}

In addition to the classification analysis, we study how the AE representation manifold correlate with the different task presented to the students. From Figure~\ref{fig:no_seg_manifold}-right, it can be seen that the AE trained with precision matrices and 20Hz stimuli (those with larger BA and sensitivity in classification) models a space whose first two coordinates (the output of the first two neurons of the Z-layer) are visibly linked to the reading efficiency for 3-syllable words (RE-words3, color-coded). Note that the smaller values of RE-words3 (associated with dyslexics) are found in the lower right corner, while the larger values are found in the upper left corner, indicating the links between an AE representation of EEG connectivity and students' reading ability. 

In Table~\ref{tab:regression_temporal}, we present a list of the tasks whose scores are significantly correlated ($p<0.1$) with the AE representation, including the Pearson's $r$ and corresponding $p$-value. Highest correlations are achieved with reading efficiency (RE), reading comprehension (RC) and verbal memory, specifically with rapid symbolic naming (RSN). The highest correlations are achieved for the precision matrix under 2Hz stimuli.
 
 \begin{table}
 	\caption{Highest Pearson's correlations ($r$ and $p$-value) between temporal connectivity matrices and assessment categories.} \label{tab:regression_temporal}
 	\begin{tabular}{p{2.2cm}rp{1.6cm}rr}
 			\toprule
 			Assessment & $f_m$ &       connectivity &         $r$ &           $p$ \\
 			\midrule
 			RE-pseudo3$^1$  &   2 &            precision &   0.569 &  $<0.001$ \\
 			RSN-objects &  2  & PC &  0.412 &   0.004 \\
 			RSN-rep-obj$^2$ &   2 & PC &  0.382 &   0.008 \\
 			RSN-objects &   2 &            precision &    0.380 &   0.008 \\
 			RC          &   2 &          correlation &  0.379 &   0.009 \\
 			\midrule
 			RC          &   8 &           covariance &  0.435 &   0.003 \\
 			\midrule
 			RE-pseudo3$^1$  &  20 &            precision &  0.433 &   0.002 \\
 			RC          &  20 &           covariance &  0.394 &   0.006 \\
 			\bottomrule
 			\multicolumn{5}{p{6.5cm}}{$^1$ 3-syllable words. $^2$ Repeated objects.}
 	\end{tabular}
 \end{table}

\subsection{Spectral Connectivity Model}\label{sec:spectral}
In the following section we analyze the results of the spectral connectivity model. Spectral connectivity measures how the power at different regions co-varies over frequency. Two different spectrum estimation methods have been used here: the direct periodogram (`pgram') and the Welch's periodogram estimation (`welch', see Sec.~\ref{sec:spectrum}). 

\begin{table*}[htp]
	\caption{Performance values for the spectral connectivity in DD diagnosis.} \label{tab:overview}
	\begin{tabular}{lrrrrrrrrrr}
			\toprule
			measure              & $f_m$ &    acc. [STD] &    sens [STD] &   spec. [STD] &    F1 & BA. &   AUC &  \\
			\midrule
			\multirow{3}{*}{\minitab[c]{precision\\(per-segment)}}     &     2 & \textbf{0.762 [0.090]} & 0.559 [0.253] & \textbf{0.835 [0.245]} & 0.649 &    \textbf{0.697} & 0.736 &  \\
			&     8 & 0.708 [0.154] & 0.543 [0.284] & 0.765 [0.251] & 0.611 &    0.654 & 0.696 &  \\
			&    20 & 0.740 [0.114] & \textbf{0.596 [0.254]} & 0.790 [0.221] & \textbf{0.660} &    0.693 & \textbf{0.762} &  \\
			\midrule
			\multirow{3}{*}{\minitab[c]{precision\\(per-subject)}}           &     2 & \textbf{0.673 [0.068]} & \textbf{0.714 [0.049]} & 0.657 [0.209] & \textbf{0.694} &    \textbf{0.686} & \textbf{0.692} &  \\
			&     8 & 0.660 [0.072] & 0.615 [0.097] & 0.676 [0.237] & 0.635 &    0.646 & 0.690 &  \\
			&    20 & 0.633 [0.161] & 0.500 [0.200] & \textbf{0.686 [0.247]} & 0.551 &    0.593 & 0.616 &  \\
			\midrule
			\multirow{3}{*}{\minitab[c]{partial\\correlation\\(per-segment)}} &     2 & 0.618 [0.135] & 0.491 [0.269] & 0.663 [0.257] & 0.537 &    0.577 & 0.560 &  \\
			&     8 & 0.694 [0.099] & 0.267 [0.362] & \textbf{0.840 [0.275]} & 0.375 &    0.554 & 0.605 &  \\
			&    20 & \textbf{0.750 [0.083]} & \textbf{0.493 [0.278]} & 0.839 [0.214] & \textbf{0.597} &    \textbf{0.666} & \textbf{0.732} &  \\
			\midrule
			\multirow{3}{*}{\minitab[c]{partial\\correlation\\(per-subject)}} &     2 & 0.571 [0.109] & \textbf{0.714 [0.104]} & 0.514 [0.281] & \textbf{0.649} &    \textbf{0.614} & \textbf{0.592} &  \\
			&     8 & \textbf{0.596 [0.163]} &   0.308 [0.363] &   \textbf{0.706 [0.324]} & 0.384 &    0.507 & 0.517 &  \\
			&    20 & 0.592 [0.121] & 0.429 [0.163] & 0.657 [0.228] & 0.484 &    0.543 & 0.590 &  \\
			\bottomrule                                 	\end{tabular}
\end{table*}

Figure~\ref{fig:seg_grouped} displays the balanced accuracy and standard deviation of the different connectivity measures, grouped by modulation frequency. Note that, as in the previous case, the values for each measure are displaced on the x-axes for a better visualization. The first comparison is between the direct periodogram and Welch's periodogram estimation methods (left and right column respectively). Here we observe that the trends are similar, except for the per-segment covariance and the per-subject partial correlation. The remaining measures behave similarly across estimation methods and approach. 

Second, we observe an important decrease in performance when changing to the per-subject approach. There is an evident direct cause: sample size. We move from 1462 segments to 49 subjects, so both the autoencoder and the SVC will be affected by less available samples during training. The least-affected measure is precision, according to Fig.~\ref{fig:seg_grouped}, being the only one that maintains a balanced accuracy around 0.7 when switching to per-subject connectivity under the Welch's periodogram estimation. 

In general, the trends of correlation and covariance are opposed to precision and partial correlation. In the per-segment approach, the former achieved better results with 2Hz modulated stimuli, whereas the later did at 20Hz. In the per-subject approach, however, the behavior was the opposite, but in almost every case, the EEG signals acquired at 8Hz (syllabic rhythm for Spanish) modulation were the worst predictor of DD. We will discuss this and its implications later in Section~\ref{sec:discussion}. To take a deeper look at the performance, we focus on precision and PC, the second best connectivity measure, shown at Table~\ref{tab:overview}.

Table~\ref{tab:overview} first confirms that the per-segment connectivity always achieves larger accuracy and sensitivity than the per-subject approach. Second, we observe that the measures related to classifier positives (sensitivity, F1 and AUC) consistently point to precision as the best-scoring adjacency measure. In the per-subject approach, sensitivity and F1 are consistently higher with the precision matrices derived from EEG segments of subjects listening to 2Hz modulated stimuli, and so are the computed AUCs, which is consistent with previous works\cite{martinez-murciaPeriodogramConnectivityEEG2019}.

\begin{figure}
	\begin{center}
		\includegraphics[width=\columnwidth]{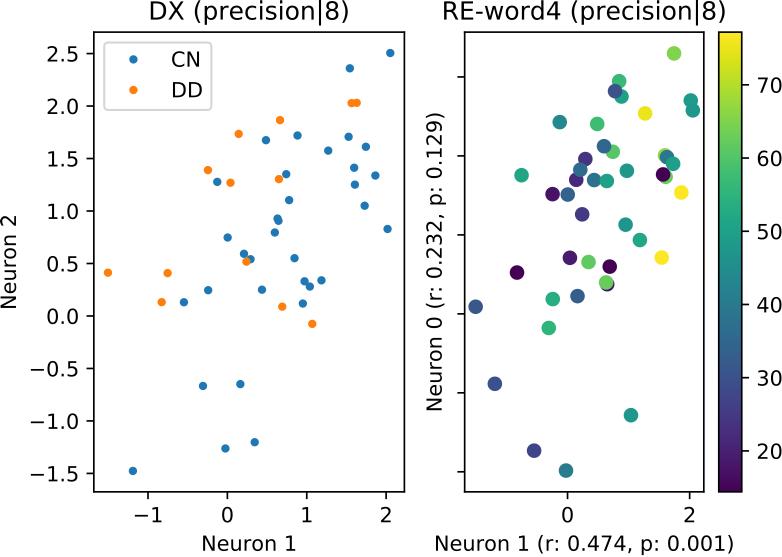}
		\caption{Manifold representation of the AE coordinate output, trained with spectral precision matrices (welch) obtained with 8Hz stimuli and the per-subject approach, and its relation with the labels (left, DD and controls) and one of the tasks (right, 4-syllable words RE).}
		\label{fig:welch_pavg_manifold}
	\end{center}
\end{figure}

Figure~\ref{fig:welch_pavg_manifold} shows the different subjects as they are modeled by the AE trained with precision connectivity matrices (welch-estimation per-subject) at 8Hz stimuli. We observe that the distribution of subjects is more disperse than in the temporal connectivity, although there are still regions where each diagnostic category predominates. As for the correlation analysis, note that the low dimensional manifold is more clearly linked to some reading assessments, such as RE for 4-syllable words, with $r>0.47$ (Fig.~\ref{fig:welch_pavg_manifold}-right).

\begin{table}
	\caption{Highest Pearson's correlations ($r$ and $p$-value) between spectral connectivity matrices (welch) and assessment categories for the per-subject scenario.}\label{tab:regression_spectral}
	\begin{tabular}{p{2.2cm}rp{1.6cm}rr}
			\toprule
			Assessment &  $f_m$ &                 kind &         $r$ &            $p$ \\
			\midrule
			EF$^1$-SA$^2$  &   2 &           covariance &  0.479 &  $<0.001$ \\
			EF$^1$-Inhibition & 2 &           covariance &  0.423 &   0.003 \\
			RS-prosody &   2 &  PC &  0.422 &   0.002 \\
			\midrule
			RE-CS$^3$ & 8 &           precision &  0.489 &  $<0.001$ \\
			RE-words4  &   8 &            precision &   0.473 &   0.001 \\
			RE-words3  &   8 &          correlation &  0.470 &   0.001 \\
			RE-pseudo4 &   8 &          correlation &  0.468 &   0.001 \\
			RE-pseudo2  &   8 &          correlation &  0.446 &   0.002 \\
			RS &   8 &          correlation &  0.424 &   0.003 \\
			RE-words3  &   8 &            precision &  0.422 &   0.004 \\
			RS  			&  20 &   precision &  0.379 &  0.007\\
			Orthography				&  20 &    PC  &  0.373 &  0.008\\
			\bottomrule
			\multicolumn{5}{p{6.5cm}}{$^1$ Executive Function. $^2$ Sustained attention. $^3$ Complex syllables}
	\end{tabular}
\end{table}

More details of the correlation analysis, under the per-subject approach, can be found at Table~\ref{tab:regression_spectral}. We first observe that the correlations are generally higher than those achieved with the temporal connectivity. The highest correlation is obtained one more time with the precision adjacency matrix, although this time with the 8Hz stimuli. Again, some assessments are highlighted: the reading efficiency (for words and pseudo-words) and reading speed. Note that a high correlation with executive functions appear with stimuli 2Hz and the PC and covariance matrices. We will discuss this later. In any case, RS and RE are repeated in both temporal and spectral connectivity analyses, hinting at possible links between EEG connectivity when listening to AM modulated noise and the ability to read in 7-y.o. students. 

\section{Discussion}\label{sec:discussion}
The main purpose of this work is to study whether and how temporal and spectral connectivity between EEG channels are linked to DD. For this purpose, we model the distribution of the connectivity matrices in high dimension using a denoising autoencoder; the encoder part of the DAE can then be used to project the adjacency --connectivity-- matrices to just a few coordinates over a collinear manifold that may be representative of each data point\cite{vincent2010stacked}. This three-dimensional space allows for an easier study of the distribution of subjects via correlation and classification analysis. 

\begin{figure}
	\begin{center}
		\includegraphics[width=0.4\textwidth]{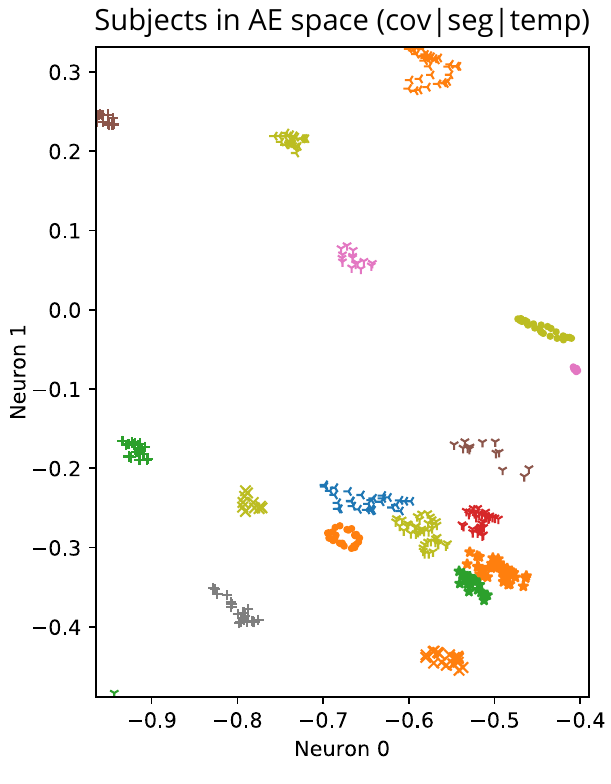}
		\caption{DAE decomposition of the temporal (temp) covariance (cov) connectivity for 2Hz stimuli under the per-segment (seg) approach. Only the first two coordinates (the output of the first two neurons of the Z layer) are shown. Points with the same color belong to the same subject.}
		\label{fig:SID_cov_seg_no_2}
	\end{center}
\end{figure}

The choice of a DAE is not trivial. In contrast to regular AEs, the addition of noise at the input makes the model learn how to separate useful information from noise. It is roughly comparable to regularization, allowing the system not to fall to local minima, at the same time that re-generating the noise in each iteration performs a moderate data augmentation. This has been consistently reported in many works\cite{vincent2010stacked,Ortiz2017,martinez-murciaStudyingManifoldStructure2019}, but in order to assess whether information is lost in the procedure, we train the DAE using the adjacency matrices from all subjects. The resulting manifold is displayed at Figure~\ref{fig:SID_cov_seg_no_2}. Specifically, we use the temporal covariance matrices acquired with the 2Hz modulated stimuli. 

In Figure~\ref{fig:SID_cov_seg_no_2}, the adjacency matrices belonging to the same subject are presented in the same color and marker. It can be easily noticed that segments from the same subject tend to cluster together in the dataset manifold. This has two major implications. First, this indicates that the adjacency matrices obtained under the same conditions (stimuli, connectivity measure, etc.) are very similar within a subject, and relatively different between subjects. Second, that the inter- and intra-subject similarities are kept when using the DAE model, and therefore it is robust to noise in adjacency matrices.

Once the DAE has been trained, it is important to address the main questions of this work. Is the DAE representation of EEG adjacency matrices related to DD? The existence of links between EEG data and dyslexia has even been doubted for some time in the literature\cite{boddyEvokedPotentialsDynamics1981,EEGReadingDisability}, and it is just very recently that EEG is starting to gain ground in the field\cite{Flanagan2018,Diliberto2018}. Conversely, the TSF for dyslexia \cite{kimppaImpairedNeuralMechanism2018,goswamiNeuralOscillationsPerspective2019, goswamiSpeechRhythmLanguage2019} states that atypical oscillatory sampling at one or more frequencies related to speech (prosody, syllable and phoneme level) could cause phonological difficulties for identifying language units, and they were shown to leave traces\cite{kimppaImpairedNeuralMechanism2018} in EEG signals.

Our results follow that line, under the assumption that an atypical oscillatory sampling may be reflected by differences in spectral connectivity when listening to AM modulated noise. Both the correlation and classification analyses showed important links between the connectivity measures and dyslexia. The per-subject DAE representation of adjacency matrices achieved correlations around 0.5 ($p<0.001$) for many dyslexia-related measures such as reading efficiency, reading speed and, in the case of spectral connectivity, executive functions. A visual inspection of the matrix representation in the DAE-space (figures~\ref{fig:no_seg_manifold} and \ref{fig:welch_pavg_manifold}) revealed consistent similarity patterns in DD and CN subjects, especially with the temporal connectivity model. For its part, the representation (color encoded) of reading efficiency measures in two of the three neurons in the Z-layer of the DAE was both visual and statistically significant (p=0.004).

The prevalence of high correlations between EEG-derived features and the assessment of phonological deficit --reading efficiency or rapid symbolic naming-- is coherent with the phonological theory of dyslexia. This theory postulates that dyslexics have a specific impairment in the representation, storage and/or retrieval of speech sounds, and the TSF found an impaired oscillatory sampling in fMRI in children and adults with dyslexia across languages and orthographies\cite{clark2014neuroanatomical}. Di Liberto et al.\cite{Diliberto2018} conducted a similar study to LEEDUCA, reporting significant correlations with similar categories, as well as other that have not been studied in this work (e.g., a test in which children had to remember whole sentences). Table~\ref{tab:comparison} compares our results to those at Ref.\cite{Diliberto2018}.

\begin{table}
	\caption{Correlation metrics of the DAE representation of EEG connectivity and the method at DiLiberto et al.\cite{Diliberto2018}. See abbreviations at Sec.~\ref{sec:leeduca}.}\label{tab:comparison}
	\begin{tabular}{lrrrr}
			& \multicolumn{2}{c}{Di Liberto\cite{Diliberto2018}} & \multicolumn{2}{c}{DAE+conn.}   \\
			Psychometric test   &    $r$ &      $p$ &   $r$ &   $p$ \\
			\midrule
			PA      			&   0.31 &    0.006 & 0.426 & 0.002 \\
			PM                  &   0.30 &    0.011 & 0.376 & 0.007 \\
			RSN                 &   0.15 &    0.200 & 0.412 & 0.003 \\
			Digit span          &   0.41 & $<$0.001 & 0.413 & 0.003 \\
			RE-words            &   0.16 &    0.179 & 0.387 & 0.007 \\
			RE-pseudo           &   0.19 &    0.099 & 0.406 & 0.004 \\
			\bottomrule
	\end{tabular}
\end{table}

\begin{figure*}[htp]
	\begin{center}
		\includegraphics[width=\textwidth]{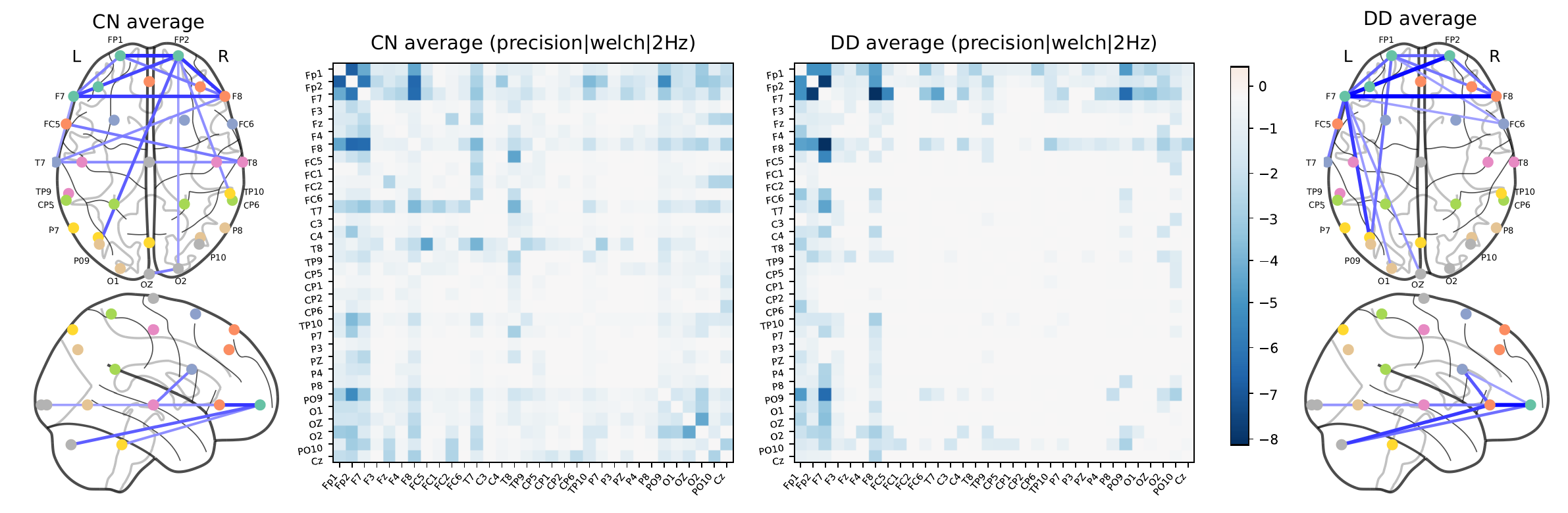}
		\caption{Average inter-channel adjacency matrices and its spatial representation over a brain template for controls (left) and dyslexic readers (right), measured by precision --inverse covariance-- over Welch's periodogram. The strength of the connection is encoded in intensity in both matrices and figures, and that the electrode placing is approximate.}
		\label{fig:connectivity}
	\end{center}
\end{figure*}

Table~\ref{tab:comparison} presents the highest $r$ between the DAE representation and the different categories, selected among all temporal and spectral connectivity measures. Observe that a similar performance is achieved for the best-performing test (digit span), whereas larger correlation values are obtained for the remaining categories, including all tests related to the phonological hypothesis (PA, RSN, PM) and reading efficiency. Correlations between the DAE representation of EEG connectivity were significant for all categories ($p<0.01$), and highly significant for PA, RSN, RE-pseudowords and digit span ($p<0.005$). This proves that there are differences in EEG connectivity between control subjects and subjects affected by DD when listening to AM-modulated white noise, and that the DAE representation of the EEG adjacency matrices is representative of the variability of DD. 

On the other hand, the classification analysis reported moderate classification performance when applying a SVC with RBF kernel on the DAE decomposition. We obtained general accuracy over 0.8 and similar BA for both spectral (0.762 and 0.740) and temporal (0.762 and 0.740) adjacency measures. It may seem moderate, but it is similar to results of recent studies using EEG to detect dyslexia\cite{perera2016review,ortiz2019anomaly}. Sensitivity was also above 0.7 in many cases and the AUC endorsed precision as the best connectivity value. 

In order to visualize the connectivity patterns that lead to a better decomposition, we show the average spectral precision matrices for groups DD and CN obtained with the 2Hz stimuli (Figure~\ref{fig:connectivity}). Note that EEG captures electric fields and not individual neural activity, and therefore electrodes placed together can capture portions of each other's signal. Here we can see obvious differences between the groups. There is strong adjacency between signals in the frontal lobe (FP1, FP2, F7 and F8), with a significantly stronger bilateral link (FP1-FP2, F7-F8) in the DD group. Note that Broca's area --a region frequently associated with language\cite{EEGReadingDisability}-- is located near F7. A relevant link between temporal regions is the T7-T8, present in CN and absent in DD. The reception field of these electrodes overlaps mainly with the primary auditory cortex. An interaction between hemispheres at multiple levels associated to phonological and prosodic processing has been reported in some fMRI studies\cite{gavrilescu2010reduced}. It would not be far-fetched to consider a reduced synchrony between left and right auditory regions as a trace of phonological deficit in subjects affected by DD, but it falls beyond the scope of this work. There are also strong connections between electrodes at the front and at the back of the head, although they are unilateral in the DD group (F7-P3, FP1-P3 and, to a lesser extent, F7-O1 and F7-OZ) and bilateral in the CN group (FP2-P3). The connection strength at the occipital region (between O1, O2 and OZ) is also higher in CN than in DD. 

In summary, this work revealed significant correlations ($p<0.005$) between a subject's performance in language tasks and composite features (DAE representation) of EEG connectivity acquired when listening to an AM modulated stimuli. Highest correlations were found with the inverse covariance matrix which yielded differences in connectivity patterns between the CN and the DD group, hinting at differences in auditory processing of speech rhythm, possibly related to the proposed atypical oscillatory sampling of the TSF. The geometric interpretation of the DAE latent space (the output of the Z-layer) allows for a visual inspection of the underlying, low-dimensional manifold, at the same time that yields a representation of the connectivity that can be used to diagnose DD with more than 80\% accuracy. The study has two major limitations: the EEG cohort is small ($n=48$, with just 16 dyslexic readers) and it is geographically limited to southern Spain. The performance difference between the per-subject ($n=48$) and the per-segment ($n=1462$) scenarios shows that the analyses could largely benefit from more data available from this and other studies. Although the results are promising and show great potential for EEG application in DD, we must be cautious in our conclusions. However, if the presence of similar EEG connectivity patterns is confirmed in earlier years (e.g., at age 4, 5 or even 6) in similar experiments, the DAE representation could be potentially used to perform an early screening of DD before the subjects have the ability to read, allowing specialized interventions for teaching reading. 

\section{Conclusions}
The main purpose of this work is to check if there are differences in Electroencephalography (EEG) connectivity between individuals affected with Developmental Dyslexia (DD) and controls, and how they are related to children's performance in different language and cognitive tasks commonly used to detect dyslexia. To do so, the manifold decomposition of a denoising autoencoder (DAE) is studied when trained with temporal and spectral EEG adjacency matrices. The resulting features inform a new low-dimensional space in which correlation and classification analysis was applied. Our results show that the DAE representation was relevant for detecting dyslexic subjects with an accuracy higher than 0.8, and a balanced accuracy around 0.7. Furthermore, the correlation $r$ between the DAE features and the language and cognitive tasks was higher than 0.5, with $p<0.005$ in many cases. We obtained higher $r$ with tasks of the phonological hypothesis category such as phonological awareness and rapid symbolic naming, as well as reading efficiency and comprehension. It is interesting to note that spectral connectivity also showed significant correlation ($p<0.001$) with measures of an executive function like sustained attention and inhibition. The precision --inverse covariance-- adjacency matrix revealed a reduced bilateral connection between electrodes of the temporal lobe (probably the primary auditory cortex) in DD subjects, as well as an increased connectivity of the F7 electrode, placed roughly on Broca's area, involved in language processing. Despite the study is geographically limited to southern Spain and the sample size is small ($n=48$), the results revealed significant links between language task performance and EEG connectivity, as well as potential to detect DD subjects using EEG signal. 

\section*{Acknowledgments}
This work was partly supported by the MINECO/FEDER under PGC2018-098813-B-C32 and RTI2018-098913-B-I00 projects. We gratefully acknowledge the support of NVIDIA Corporation with the donation of one of the GPUs used for this research. Work by F.J.M.M. was supported by the MICINN ``Juan de la Cierva - Formaci\'on'' FJCI-2017-33022. We also thank the \textit{Leeduca} research group and Junta de Andaluc\'ia for the data supplied and the support.

%
%

\end{document}